\documentclass[10pt]{article}
\usepackage{CustomP,amssymb,xcolor}

\newcommand{\z}{\zeta}

\title{A Hamiltonian Bifurcation Perspective on Two Interacting Vortex Pairs: From
Symmetric to Asymmetric Leapfrogging, Period Doubling and Chaos}

\author[1]{Brandon Whitchurch}
\author[1]{Panayotis Kevrekidis}
\author[2,1,3]{Vassilis Koukouloyannis}

\affil[1]{Department of Mathematics and Statistics, University of Massachusetts, Amherst, MA 01003-4515, USA}
\affil[2]{Department of Mathematics, Statistics and Physics, College of Arts and Sciences, Qatar University, P.O. Box 2713, Doha, Qatar}
\affil[3]{Faculty of Civil Engineering, School of Engineering, Aristotle University of Thessaloniki, 54249, Thessaloniki, Greece}
\begin{document}
\maketitle
\abstract{In this work we study the dynamical behavior of two interacting vortex pairs, each one of them consisting of two point vortices with
  opposite circulation in the 2d plane. The vortices are considered as effective particles and their interaction can be desribed in classical mechanics terms. We first construct a Poincar\'e section, for a typical value of the energy, in order to acquire a picture of the structure of the phase space of the system. We divide the phase space in different regions which correspond to qualitatively distinct motions and we demonstrate its different temporal evolution in the ``real'' vortex-space. Our main emphasis is on the leapfrogging periodic orbit,
  around which we identify a region that we term the ``leapfrogging envelope''
  which involves mostly regular motions, such as higher order periodic and quasi-periodic solutions.  We also identify the chaotic region of the phase plane surrounding the leapfrogging envelope as well as the so-called  walkabout and braiding motions. Varying the
  energy as our control parameter, 
  we construct a bifurcation tree of the main leapfrogging solution and its instabilities, as well as the instabilities of its daughter
  branches.
  We identify the symmetry-breaking instability of the
  leapfrogging solution (in line with earlier works), and also
  obtain the corresponding asymmetric branches of periodic solutions.
  We then characterize their own instabilities (including period
  doubling ones) and bifurcations in an effort to provide a more
  systematic perspective towards the types of motions available to this
  dynamical system. }

\section{Introduction}

In a fundamental work published in 1893, Love~\cite{Love}
analyzed the motion of two vortex pairs with a common axis.
His remarkable analysis showed that two such vortex pairs
(consisting of dipoles, i.e., vortices with opposite
charges) will perform a leapfrogging motion around each
other if the ratio of their diameters, at the precise moment one passes through the other, was larger than
a critical one $\alpha_1=0.172$. His motivation was
to emulate a simpler, more tractable 2d analogue of
the 3d setting of vortex rings (VRs), earlier analyzed
by Helmholtz.  For $\alpha < \alpha_1$, the smaller,
faster ring manages to escape the velocity field imposed by the
wider, slower ring and the two separate indefinitely.
The story lay dormant for a considerable while until the
use of computers gave Acheson~\cite{Acheson} the ability to
explore the relevant phenomenology numerically. This resulted
in the identification of a second critical point $\alpha_2=\phi^2=0.382$
where $\phi=(\sqrt{5}-1)/2$ is the golden ratio below which
the leapfrogging motion turns out to be unstable.
In fact, Acheson~\cite{Acheson} claimed the existence of two different
instability modes, one for $0.172< \alpha <0.29$ and
another one for $0.29 < \alpha < 0.382$. For $\alpha>0.382$ the leapfrogging solution is linearly stable. A careful more
recent analysis on the basis of Floquet theory has led
Toph\o j and Aref~\cite{Tophoj} to conclude also that the actual leapfrogging
instability occurs at the square of the golden ratio.

These works are of considerable interest due to the intrinsically
complex yet tractable nature of the vortex leapfrogging
problem. In addition, recent works both at the level of analysis
of experiments performed in liquid helium~\cite{wacks} and at
that of exploring vortex rings in confined atomic Bose-Einstein
condensates~\cite{komineas} (and also recent works such
as~\cite{caplan,wang1,wang2}) render this phenomenology
quite interesting to understand and generalize. In the context of
superfluid helium experiments, there is an interest in exploring
generalized leapfrogging of three, or more vortex rings~\cite{wacks}.
On the other hand, in the BEC realm incorporating the role of the trap
(which will destroy leapfrogging but will generate coherent in-trap
motions of vortices and
VRs~\cite{wang1,wang2}) is a subject both theoretically
relevant and experimentally accessible. 

Our aim in the present work is to provide some perspective
on the full realm of possible motions and a clearer picture of the
bifurcation structure of the vortex pair motion from the point
of view of particle mechanics. Since the system possesses a Hamiltonian structure, the appropriate tool to perform such a study is to construct the Poincar\'e
surface of section (PSS) associated with the vortex motion, in order
to appreciate and separate the regions of different accessible
motions to the system. {While our considerations
  are chiefly focused on the central (to
  the leapfrogging motion) periodic orbit and its instability threshold,
  we also examine
the orbits stemming from this
symmetry breaking and further bifurcations to which these states
become subject to as the energy is further decreased}. The resulting
bifurcation diagram provides a more global perspective on the possible
orbits, stability and associated bifurcations in the system, while
the PSSs offer some perspective on the potential chaoticity of the
system of 4 vortices and how this emerges within the dynamics.
In that sense our work corroborates and extends earlier findings
of the classic papers of~\cite{{Love},{Acheson}} and the more recent analysis
of~\cite{Tophoj} and sheds some light on techniques that may be
useful when exploring generalizations of leapfrogging such as the
ones under consideration in superfluid helium.

The manuscript is organized as follows. In section \ref{system} we
discuss the system of interest. We begin by revisiting the general equations describing the interaction of point vortices considered as effective
particles. Since we consider two vortex pairs in our system, we can use the
integrals of motion to suitably reduce the dynamics and thus, to end up with a two degrees of freedom setting. In Section \ref{section} a proper PSS is defined. In section \ref{regions} we acquire a PSS for a typical value of the energy and describe the various areas in this section which correspond to qualitatively different motions. The
temporal evolution of these motions in the real $X$-$Y$ plane of motion is also described. In addition, in section \ref{bifurcations} we produce the bifurcation
tree of the central leapfrogging solution (upon its destabilization)
with the total energy of the system as a parameter.
We thus characterize not only the standard and more well known leapfrogging
motion, but also its asymmetric variants, and their own bifurcations.
Relevant pitchfork and period doubling phenomena are revealed and
also the potential of the phase space to feature chaotic motion is
showcased. Finally, in section~\ref{conclusions}, we summarize our
findings and present some relevant directions for future study.

\section{The System of Two Interacting Vortices}\label{system}

In what follows we consider the vortices as effective point particles moving on a $X$-$Y$ plane, as is often done in fluid and superfluid
contexts~\cite{newton1,newton2,PGK}. The equations of motion describing the dynamical behavior of a general system of $N$ interacting vortices is given by e.g.~\cite{Pomphrey}
\begin{equation}
\overline{\frac{d z_\alpha}{d t}}=\frac{1}{2\pi i}{\sum_{\beta=1, \beta\neq \alpha}^N}\frac{k_{\beta}}{z_{\alpha}-z_{\beta}},\quad \alpha=1\ldots N
\label{eqsmot}
\end{equation}
where $z_\alpha=X_\alpha+i\, Y_\alpha$ is the position of the $\alpha$-vortex given in complex coordinates, the bar denotes complex conjugation, $k_{\alpha}$ denotes the circulation of the corresponding vortex.

The system possesses also a Hamiltonian structure, and the corresponding Hamiltonian is given by~\cite{Pomphrey,Eckhardt}
\begin{equation}
	H=-\frac{1}{2\pi} \sum_{1\leq\alpha<\beta\leq N} k_{\alpha} k_{\beta} \ln|z_{\alpha}-z_{\beta}|.
	\label{eq:genHam}
\end{equation}

In the present work we study a system consisting of two vortex pairs. Each one of the pairs consists of two vortices having opposite circulation. The corresponding complex coordinates for each vortex are  $z_1^+, z_1^-, z_2^+, z_2^-$, where the (+)-superscript denotes the vortices with the positive (counterclockwise) circulation while the (-)-superscript denotes the vortices with negative (clockwise) circulation. We consider the case of vortex pairs of the same circulation magnitude i.e. $k_{1,2}^+=1$ and $k_{1,2}^-=-1$.  

The equations of motion in this case become~\cite{Tophoj}
\begin{eqnarray}
\overline{\frac{d z_1^+}{d t}}=\frac{1}{2\pi i}\left(\frac{1}{z_1^+-z_2^+}-\frac{1}{z_1^+-z_1^-}-\frac{1}{z_1^+-z_2^-}\right)\\
\overline{\frac{d z_2^+}{d t}}=\frac{1}{2\pi i}\left(\frac{1}{z_2^+-z_1^+}-\frac{1}{z_2^+-z_1^-}-\frac{1}{z_2^+-z_2^-}\right)\\
\overline{\frac{d z_1^-}{d t}}=\frac{1}{2\pi i}\left(\frac{1}{z_1^--z_1^+}+\frac{1}{z_1^--z_2^+}-\frac{1}{z_1^--z_2^-}\right)\\
\overline{\frac{d z_2^-}{d t}}=\frac{1}{2\pi i}\left(\frac{1}{z_2^--z_1^+}+\frac{1}{z_2^--z_2^+}-\frac{1}{z_2^--z_1^-}\right)
\end{eqnarray}

In order to decrease the number of degrees of freedom considered within
the system, we perform the cannonical transformation~\cite{Eckhardt, Tophoj}

\begin{equation}
\begin{array}{lcl}
\z_0&=&\frac{1}{2}\left(z_1^++z_2^+-z_1^--z_2^-\right)\\[8pt]
\hat{\z}_0&=&\frac{1}{2}\left(z_1^++z_2^++z_1^-+z_2^-\right)\\[8pt]
\z_+&=&\frac{1}{2}\left(z_1^+-z_2^++z_1^--z_2^-\right)\\[8pt]
\z_-&=&\frac{1}{2}\left(z_1^+-z_2^+-z_1^-+z_2^-\right).\label{trans2}
\end{array}
\end{equation}

In this set of variables, $\z_+$ is the vector connecting the center of separations $d_1=z_1^+-z_1^-$ and $d_2=z_2^+-z_2^-$ of the two vortex-pairs, while $\z_-$ is the corresponding difference, $\z_-=\frac{1}{2}(d_1-d_2)$. On the other hand $\z_0=\frac{1}{2}(d_1+d_2)$ is half the linear impulse of the system  that (as confirmed  also below) constitutes an integral of motion of the system~\cite{newton1,newton2}, while, its conjugate variable $\hat{\z}_0$ stands for twice the centroid of the system.

The reduced Hamiltonian after the action of the transformation (\ref{trans2}) is
\begin{equation}
        H = -\frac{1}{2\pi}\ln\left|\frac{1}{\zeta_0^2-\zeta_+^2}-\frac{1}{\zeta_0^2-\zeta_-^2}\right|.
\end{equation}
We can see here that the variable $\hat{\z}_0$ is cyclic (it does not explicitly appear in the Hamiltonian) which also verifies that its conjugate variable $\z_0$ is an integral of motion. Having that in mind, and by a proper re-scaling, we get the following form for the Hamiltonian 
\begin{equation}
        H = -\frac{1}{2}\ln\left|\frac{1}{1+\zeta_{+}^{2}}-\frac{1}{1+\zeta_{-}^{2}}\right|.
        \label{eq:rescaledHam}
\end{equation}
We now have all that is required to obtain the equations of motion in terms of these rescaled $\zeta$ variables.
\begin{subequations}
        \begin{equation}
                \overline{\frac{d\zeta_{+}}{dt}} = i\zeta_{-}\left(\frac{1}{\zeta_{+}^{2} - \zeta_{-}^{2}} + \frac{1}{1+\zeta_{-}^{2}}\right),
        \end{equation}
        \begin{equation}
                 \overline{\frac{d\zeta_{-}}{dt}} = i\zeta_{+}\left(\frac{1}{\zeta_{-}^{2} - \zeta_{+}^{2}} + \frac{1}{1+\zeta_{+}^{2}}\right)
        \end{equation}
        \begin{equation}
                 \overline{\frac{d\hat{\zeta}_{0}}{dt}} = \frac{1}{1+\zeta_{+}^{2}} + \frac{1}{1+\zeta_{-}^{2}}.
        \end{equation}
        
				\label{eq:ReScaledEoMs}
\end{subequations}
Equation \ref{eq:ReScaledEoMs}c is the equation of motion for the system centroid at a constant speed. Although this equation is not needed for the study of leapfrogging and relevant motions, it is necessary for the reconstruction of the orbits of the vorices in the real vortex-plane.

In order to acquire the real counterparts of the equations of motion which are necessary for the numerical investigation of the system we consider the following form of $\z_+$ and $\z_-$
\begin{equation}
        \zeta_{+} = x_{1} + ix_{2}, \qquad \zeta_{-} = y_{1} + iy_{2}. \label{zeta}
\end{equation}
By substituting the above definition (\ref{zeta}) into (\ref{eq:ReScaledEoMs}) we get the required four equations with respect to the
variables $x_1, x_2, y_1$ and $y_2$\footnote{In order to avoid confusion, please note that with the capital $X_i, Y_i$ we denote the spatial coordinates of the $i$-vortex in the original space, while the small lettered $x_1, x_2, y_1, y_2$ denote the real and imaginary parts of the relative coordinates of $\z_+$ and $\z_-$.}.

\section{Numerical Investigation}

\subsection{Construction of the Poincar\'e Surface of Section} \label{section}
Upon the above transformations discussed
also in~\cite{Tophoj}, the system under investigation is transformed
into a two-degrees-of-freedom one. The standard Hamiltonian-Mechanics tool to investigate such systems and obtain a comprehensive picture of their phase-space is the Poincar\'e surface of section (PSS).

The first step in order to construct an appropriate PSS is to consider a specific value of the energy $h$ of the system which is the nominal value of the Hamiltonian for a set of initial conditions i.e. $h=H(x_{10}, x_{20}, y_{10}, y_{20})$. In what follows, the value of $h$ will be our main control parameter.

The second step in order to define the PSS is to define a section crossing by choosing a fixed value of one of the variables. Geometrically, this step defines a plane in the phase space which all\footnote{In the case where this is not possible, we choose instead a plane which a representative set of orbits intersects.} solutions should intersect. The symmetry of the main leapfrogging solution causes it to remain in the plane where $x_2=y_1=0$ i.e.~all of the motion occurs in the $x_1$ and $y_2$ directions; thus, in order to ensure that the leapfrogging solution crosses our PSS we must define the section plane to be transverse to the $x_2=y_1=0$ plane. In particular, we choose the section plane to be at $x_1=0$, which corresponds to the configurations of vortex-pairs that have their centroids aligned vertically.

\begin{figure}[h!]
\centering
\includegraphics[width = 12cm]{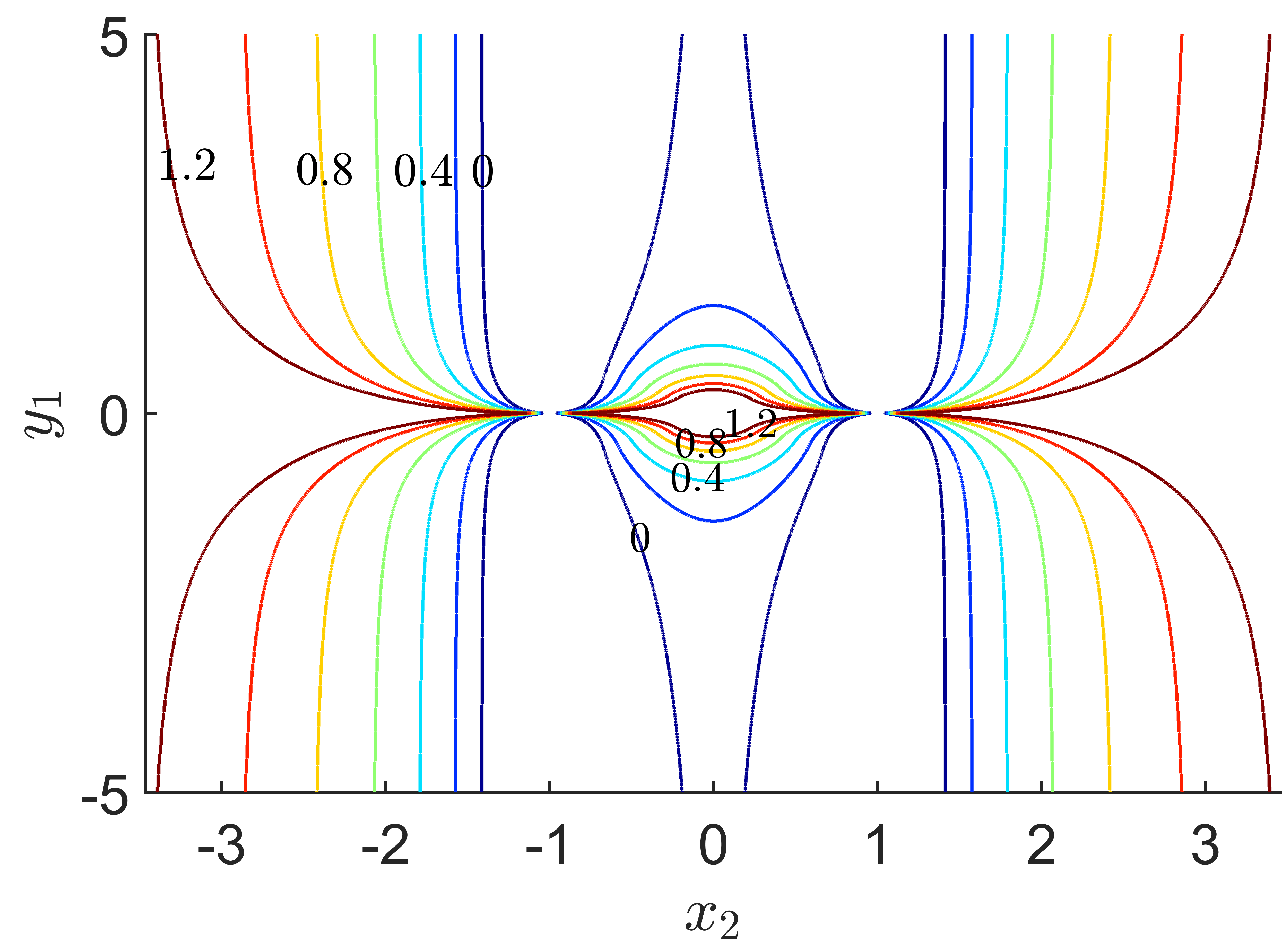}
\caption{Contours showing the boundary of the permissible area of the Poincar{\'e} section in the $x_2$-$y_1$ plane for values of $h$ ranging from $0$ to $1.2$. We can see how this area shrinks for increasing values of $h$ which are depicted as labels of the corresponding contours.}
\label{permittedContours}
\end{figure}

Each point of a PSS should uniquely define the set of initial conditions of an orbit for our system. The initial value of $x_1$ is fixed by the definition of our section plane, hence $x_{10} = 0$. {Furthermore, all of our initial conditions must satisfy the energy requirement. This means that only two of the three remaining initial values can be chosen, with the third one being determined by the fixed value of $h$. The initial values of $x_2$ and $y_1$ are the ones which we allow to vary independently (this has the convenient byproduct that the main leapfrogging solution is located at the point $(0,0)$ on our section) and thus initial values of $y_2$ must calculated using the values of $h$, $x_1$, $x_2$, and $y_1$. Then all of the initial conditions can be given as quadruplets of the form $(0, x_2, y_1, y_2(x_2, y_1, h))$}. The equation for $y_2$ is a polynomial of degree four and hence provides four solutions. In order to ensure the unique correspondence of a point in the plane to the initial conditions of an orbit we have to ensure that we always make the same selection, which in this case is the smaller positive one. This choice is called the ``direction'' of the orbit and obviously it has to provide a real value for $y_2$, 
 but, this is not always possible. There are values of $x_2, y_1$ such that a real $y_2$ does not exist for the specific value of the energy $h$ and direction. The set of $(x_2, y_1)$ points in the PSS which provides an appropriate $y_2$ is called the ``permissible'' area and coincides also with the region where the motion of our system is permitted. 
In Fig.~\ref{permittedContours} the boundary of the permissible region for various values of the parameter $h$ is shown, where we can observe how the permissible area is shrinking as the $h$ increases from $h=0$ to $h=1.2$.

After choosing a point on the section, the next point on the section is acquired when the corresponding orbit will cross $x_1=0$ with the chosen direction. This way we get a comprehensive picture of the kind of accessible orbits. A set of distinct points on the surface correspond to periodic orbits of the system,
(apparently) ``continuous'' curves correspond to quasi-periodic, while an ``unorganized cloud'' of points corresponds to chaotic motion. 

In~\cite{Acheson}, the initial conditions of the leapfrogging solution are given as
\begin{equation}
        x_{1} = 0,\  x_{2} = 0,\  y_{1} = 0,\  y_{2} = \frac{1-\alpha}{1+\alpha},
\end{equation}
where $\alpha$ is the ratio of the width of the two vortex pairs at the instance one passes through the other, and ranges between $0$ and $1$. The parameter $\alpha$ is well defined only for symmetric orbits, in order to perform a more general investigation, we choose the energy $h$ of the system to be our main study-parameter instead of $\alpha$. The relation between $\alpha$ and the energy $h$ is given by~\cite{Tophoj} 
\begin{equation}
        \frac{4\alpha}{(1-\alpha)^{2}} = e^{2h},
        \label{eq:Alpha2Ham}
\end{equation}
and can provide a connection between our works and previous literature.


\subsection{Main PSS's areas - Orbit types} \label{regions}

\begin{figure}[!h]
\centering
\begin{tabular}{cc}
				(a)&(b)\\
        \includegraphics[width=8cm, height = 7cm]{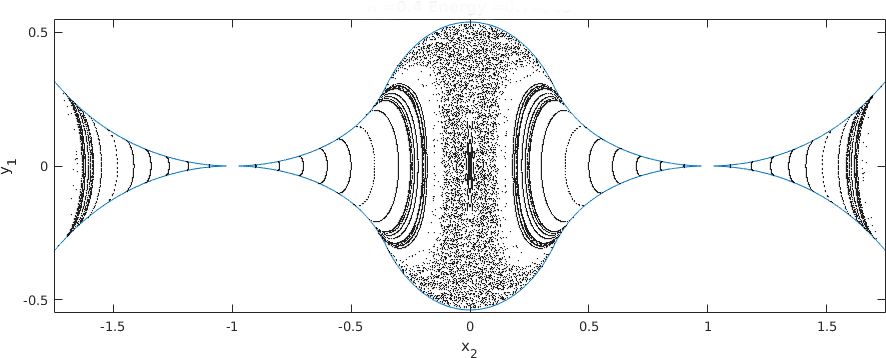}&\includegraphics[height=7cm]{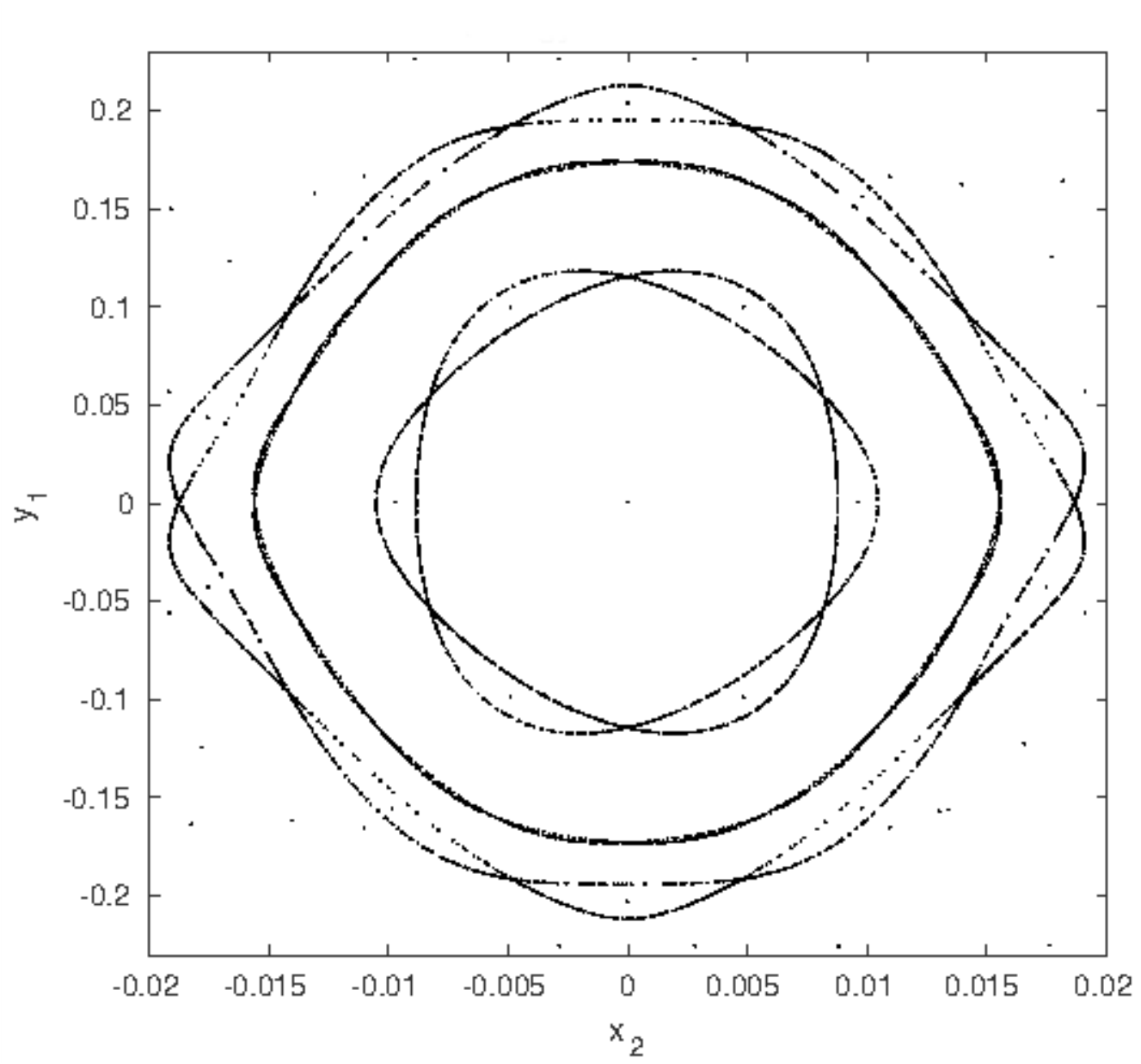}
\end{tabular}
\caption{The Poincar\'e Section for $h = 0.7458$. (a): The permissible area of motion of the system lies inside the blue curve. Inside this area we can distinguish the leapfrogging envelope around $(0,0)$ (see also (b) for a detailed zoom into that area). The leapfrogging envelope is surrounded by a chaotic region beyond
which one finds the area of walkabout motion which extends up to approximately $|x_2|<1.5$. In the outer parts of the section lies the area of the braiding motion. (b) zoom into the Leapfrogging Envelope. A close up view of the region around the $(0, 0)$ point which corresponds to the main leapfrogging motion. We can distinguish many higher order resonances around the central periodic orbit. These higher-period periodic orbits,  correspond to motions which have the same characteristics as the leapfrogging one.}
        \label{fig:Section7458}
\end{figure}

According to~\cite{{Acheson},{Love}}, the leapfrogging motion is known to be stable for parameter values greater than $\alpha > \phi^2 \simeq 0.382$. Since Poincar\'e sections are more easily obtained in the stable regime, a preliminary section was constructed for a value of $h\simeq 0.7458$ which corresponds to $\alpha = 0.4$ (Fig.~\ref{fig:Section7458}(a)). This section was used as a reference when constructing sections at other energy levels 
, and as such it is useful to characterize the various areas of the section which correspond to distinct types of motion for our system. Sections for other values of $h$ are shown in Fig.~\ref{MultiSection}.

First of all, we can see in Fig.~\ref{fig:Section7458}(a) that the section does not occupy the whole plane. Instead, as it is discussed in the previous subsection, the motion is restricted inside the permissible area which is denoted by a blue curve. Although, every point of the permissible area corresponds to a valid initial condition for an orbit we chose only some of them in order to be able to focus to specific representatives of the motions we want to study.

\begin{figure}[!h]
\centering
\begin{tabular}{cc}
				(a)&(b)\\
        \includegraphics[width = 8cm]{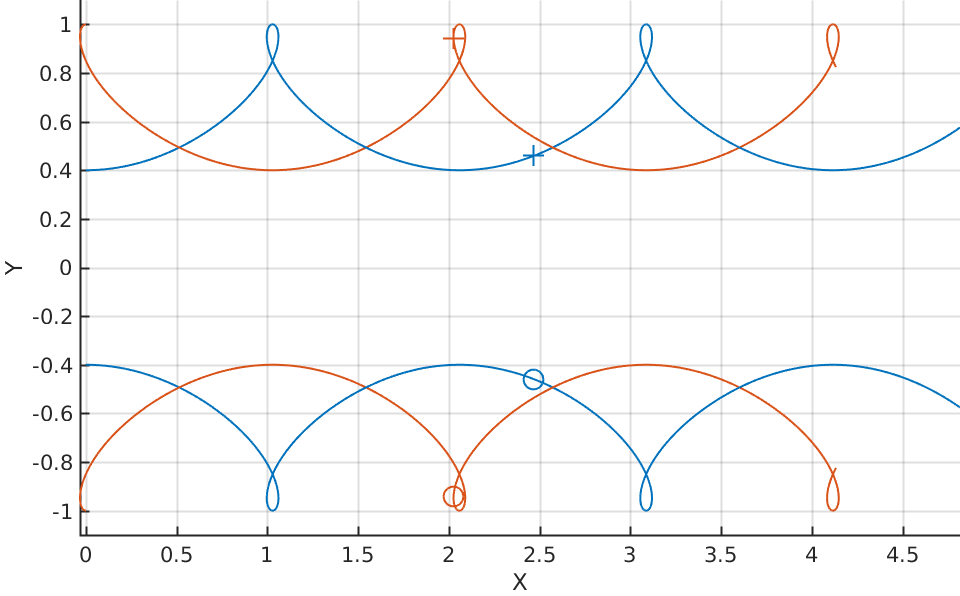}&\includegraphics[width = 8cm]{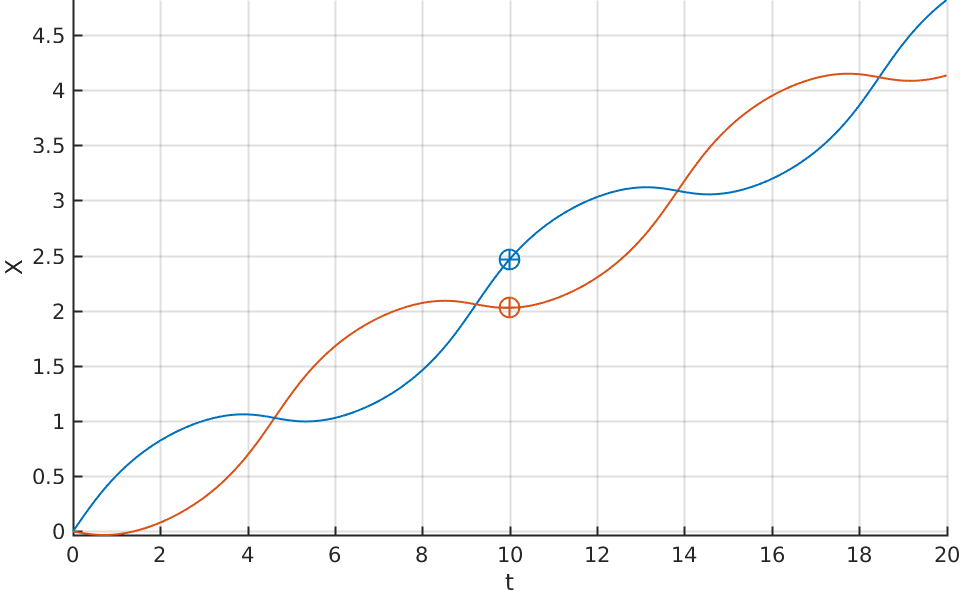}
\end{tabular}
				\caption{
The temporal evolution of the main leapfrogging motion for $h=0.7458$. Crosses are used for vortices with positive circulation; circles are used for vortices with negative circulation. (a): The motion in the vortex $X$-$Y$ plane. (b) The leapfrogging motion on the $t$-$X$ plane. We can observe that the orbit is completely symmetric in the sense that the $X$ coordinates of the vortices of the same pair coincide. Every two intersection points in (b) corresponds to one point on the Poincar\'e section.}
        \label{fig:GeneralLeap}
\end{figure}

As mentioned before, the {\it leapfrogging motion} that was analyzed by Love~\cite{Love} corresponds to the $ (0,0)$ (fixed) point of the PSS. A magnification of the area around this orbit is shown in Fig.~\ref{fig:Section7458}(b). The time evolution of this solution is depicted in Fig.~\ref{fig:GeneralLeap}\footnote{For a video description of the motions studied in this work one can refer to \cite{Videos}.}. 
In this kind of motion the vortices of the same circulation rotate around each other. In addition, the X-coordinate of the two vortices of each pair is always the same, while, when one of the vortices passes inside the other, all four vortices have the same X-coordinate value. In this sense, this motion is reminiscent of the corresponding
leapfrogging of vortex rings~\cite{caplan,konstantinov}.
The projection into the $t$-$X$ plane also makes it possible to distinguish where the section crossings occur; for this case it is wherever the paths in Fig.~\ref{fig:GeneralLeap}(b) intersect with the chosen direction. Thus we get a point in the PSS every two $X$-orbit sections. 

The area surrounding the leapfrogging solution on the PSS is shown in Fig.~\ref{fig:Section7458}(b). In this figure, besides the central leapfrogging solution which lies at $(0, 0)$, we can identify many higher-period periodic orbits surrounding the central one. These orbits are associated with
periodic motions bearing the characteristics of the leapfrogging motion but their return to the exactly same configuration occurs after more than one section crossings. Although there is this fundamental difference between the period-one central orbit and the higher-period ones, since they lie very close in the PSS, they are not visually distinguishable in the vortex-plane and hence we do not present any related figure. Given the similarity in the qualitative characteristics of such orbits as the central one, we choose to name this region the {\it leapfrogging envelope}. Note that, not all the possible orbits  of this region are shown in Fig.~\ref{fig:Section7458}(b). We choose to show some indicative ones that reveal the structure of the PSS, i.e. that this is a region mostly of regular (periodic and quasi-periodic) motion, and that there exist also higher order periodic orbits. 

The leapfrogging envelope is surrounded by a chaotic region, which extends from the leapfrogging envelope up to approximately $|x_2|\simeq0.2$. Since this region is not confined by any invariant curves, the whole region is inhabited by just one orbit.\par

\begin{figure}[!h]
\centering
\includegraphics[width=9cm]{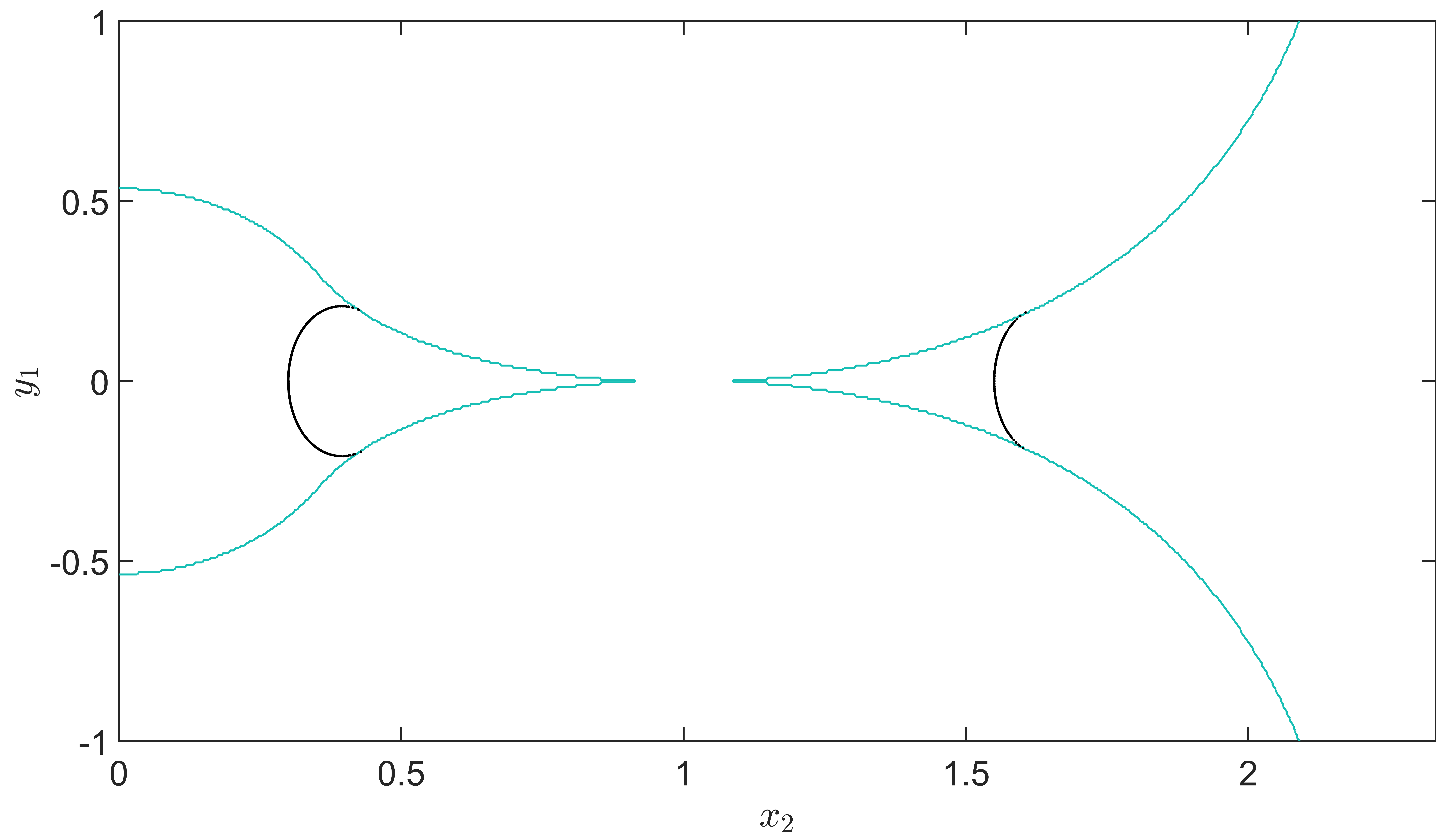}
\caption{A sigle Walkabout orbit in the PSS for $h=0.7458$. We can observe its attribute of crossing the $x_2=1$ asymptote.}
\label{SingleWalkaboutSection}
\end{figure}

Outside of the chaotic region which encloses the leapfrogging envelope there is another region of regular behavior, which lies approximately in the initial conditions region defined approximately by $y_2=0$ and $0.2<|x_2|<1.5$. The orbits that exist in this region, are the so-called {\it Walkabout Orbits}~\cite{{Acheson},{Tophoj}} and are mostly of quasi-periodic character and such would be also if we had chosen initial conditions between the depicted curves. 
\begin{figure}[!h]
\centering
\begin{tabular}{cc}
	(a)&(b)\\
        \includegraphics[width = 8cm]{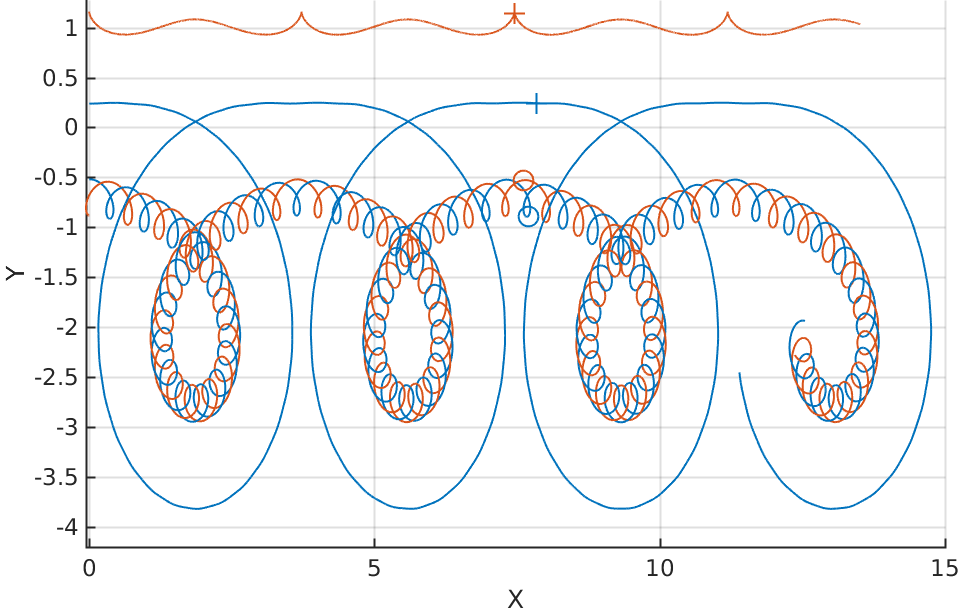}&\includegraphics[width = 8cm]{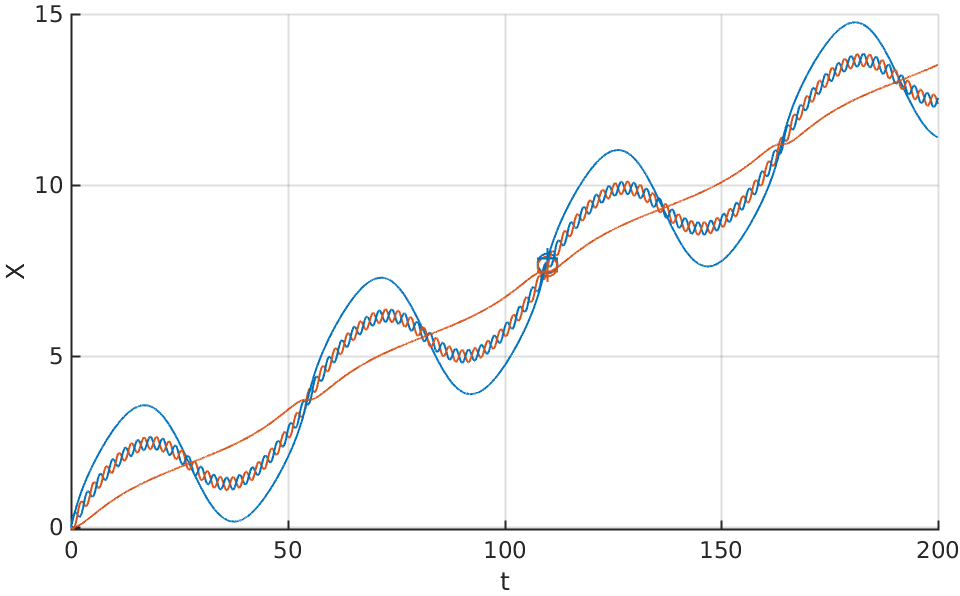}
\end{tabular}
        \caption{(a): Temporal evolution of a walkabout orbit for $h=0.7458$ in the $X$-$Y$ plane. (b): The walkabout motion in the $t$-$X$ plane.}

        \label{fig:Walkabout}
\end{figure}
A walkabout orbit is shown in Fig.~\ref{fig:Walkabout}.
 In this kind of motion two of the vortices with the same circulation rotate quickly around each other, while one of the vortices with the opposite circulation rotates more slowly around the former two. Finally the last vortex performs a motion which lies always outside of the rotation of the other three. The walkabout orbits exhibit an interesting property in that they are able to jump across the $|x_{2}|=1$ asymptote on the PSS (Fig.~\ref{SingleWalkaboutSection}). This happens because at the moment when the centroids of the two vortex-pairs are vertically aligned (and thus we have a section crossing), a vortex can lie ``inside'' the other vortex-pair (i.e. $|x_{2}| < 1$) or both vortices of the vortex-pairs with lie completely ``outside'' the other ($|x_{2}| > 1$). \par

The chaotic region, which was mentioned before, includes itself many motions which can exhibit
features (or transient fractions of the motion) reminiscent of
leapfrogging or walkabout behavior but due to their chaotic nature they do
not maintain either of these behaviors. Many of these regions behave similarly to the motions studied previously wherein the vortices leapfrog for some time before breaking into some walkabout motion, and then sometimes returning to
leapfrogging. 

\begin{figure}[!h]
\centering
\includegraphics[height = 7cm]{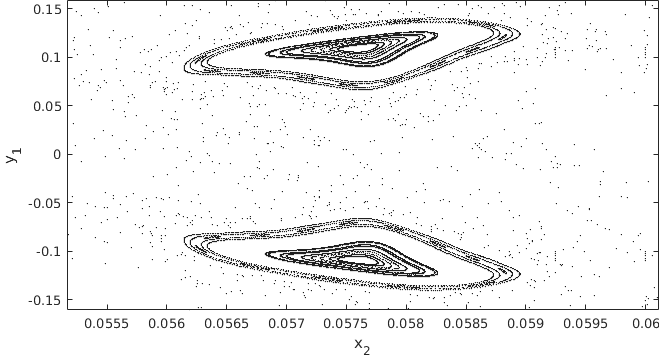}
\caption{A magnification of Fig.~\ref{fig:Section7458} around the islands corresponding to the mixed period leapfrogging motion.}
\label{MixedPerSection}
\end{figure}

\begin{figure}[!h]
\begin{tabular}{cc}
(a)&(b)\\
\includegraphics[width = 8cm]{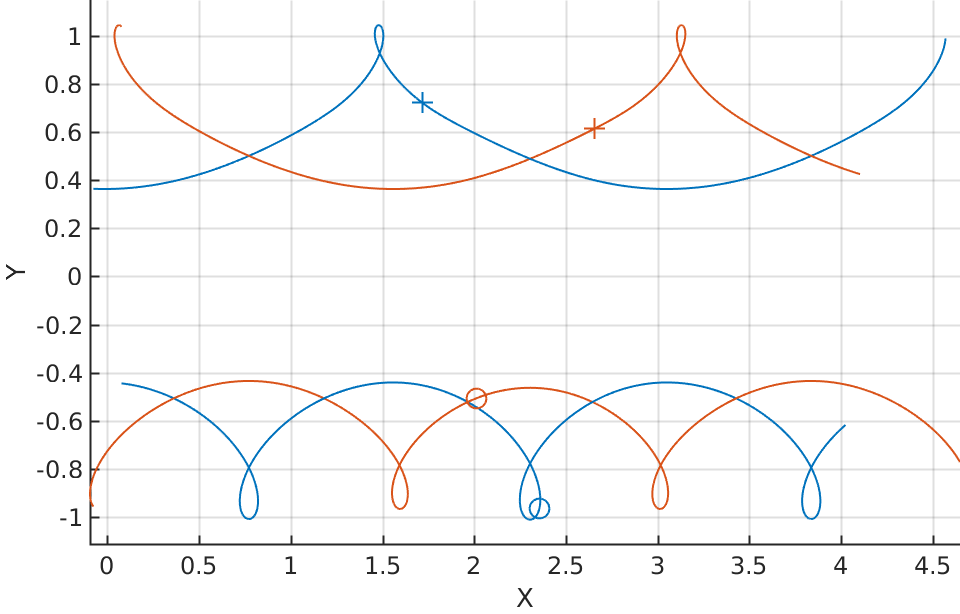} & \includegraphics[width=8cm]{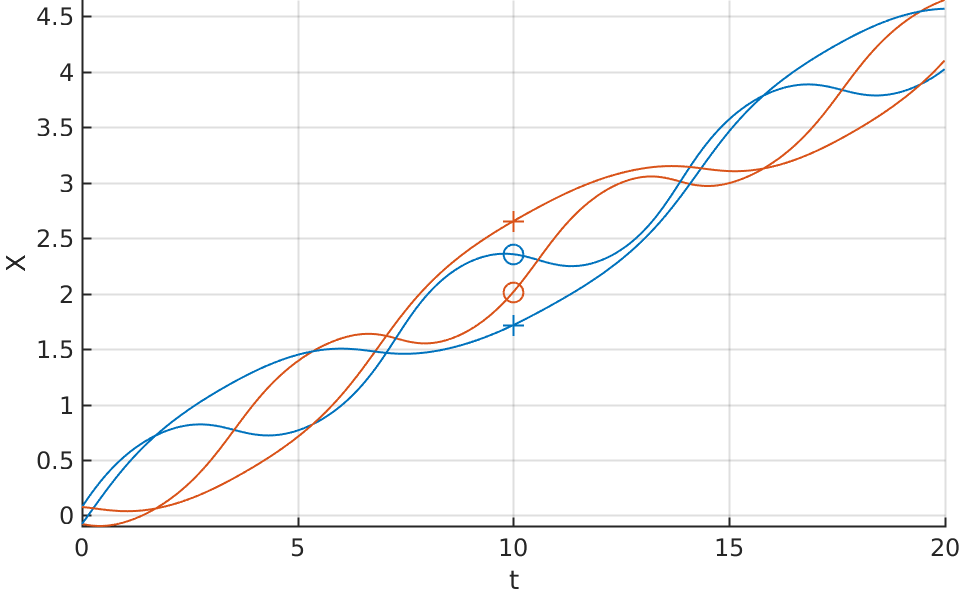}
\end{tabular}
        \caption{(a): Temporal evolution of the Mixed Period Leapfrogging motion for $h=0.7458$ in the $X$-$Y$ plane. (b): Evolution of Mixed Period Leapfrogging in the $t$-$X$ plane. This kind of motion is similar to leapfrogging, but with vortex pairs of like circulation orbiting each other at different speeds.}
        \label{fig:MixedPeriod}
\end{figure}

There is another class of orbits which thus far has only been found to exist within the chaotic region and corresponds to the islands shown in the fraction of the PSS shown in 
Fig.~\ref{MixedPerSection}. This class of
orbits known as mixed period leapfrogging and is characterized by a motion similar to leapfrogging, but with vortex pairs of like circulation orbiting each other at different speeds. Fig.~\ref{fig:MixedPeriod} shows an example of this type of motion.\par

\begin{figure}[!h]
\begin{tabular}{cc}
(a)&(b)\\
        \includegraphics[width = 8cm]{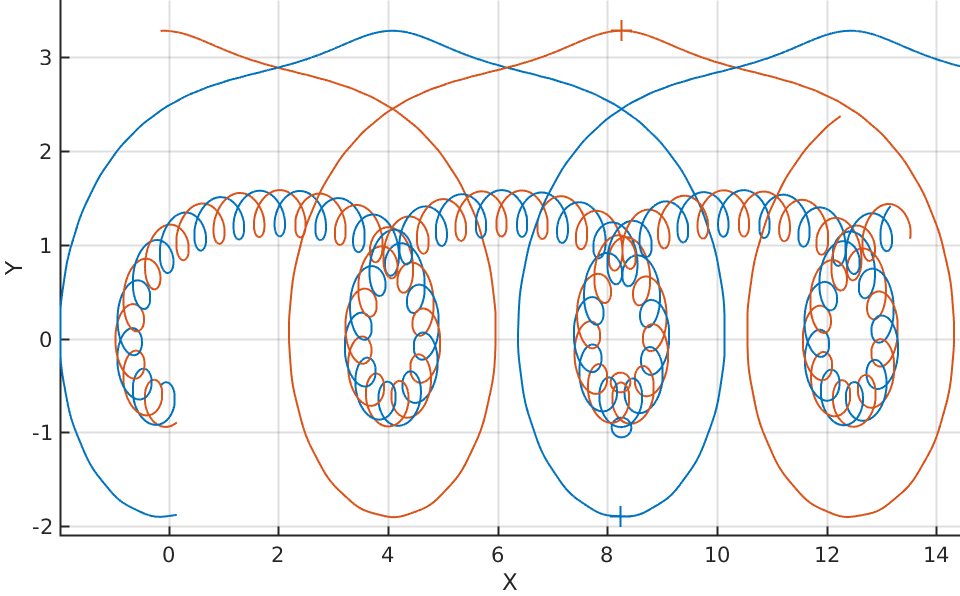}&\includegraphics[width = 8cm]{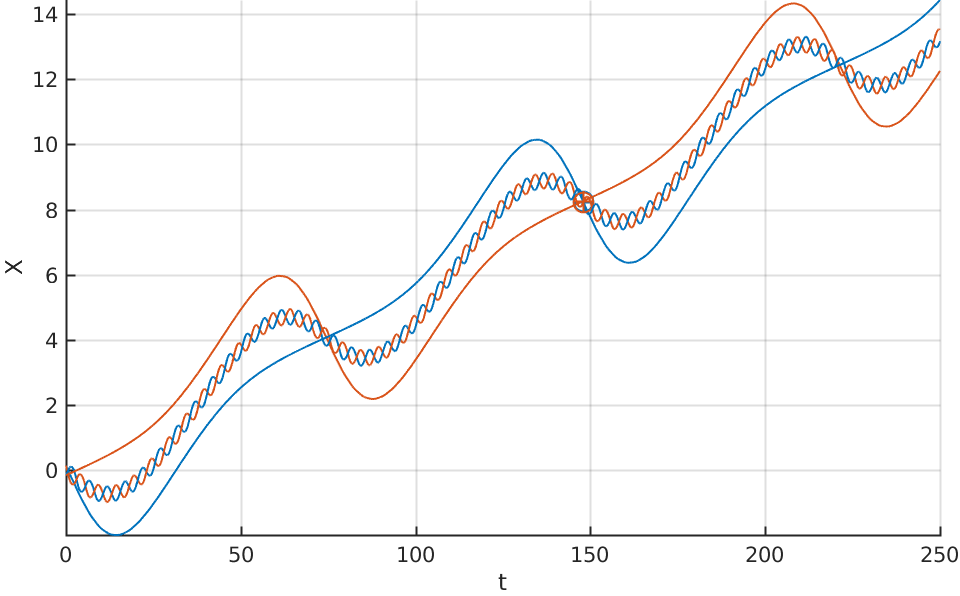}
\end{tabular}
        \caption{(a): Temporal Evolution of a Braiding Orbit for $h=0.7458$ in the $X$-$Y$ plane. (b): Evolution of the braiding orbit in the $t$-$X$ plane.} 
        \label{fig:Braiding}
\end{figure}

The final region of the section which have yet to be discussed are those regions on the outer edges of the permissible region which lie approximately for $|x_2|>1.5$.
The motion in this region (which has come to be called the {\it Braiding Motion}) has section crossings where the vortex pairs are always outside of each other, hence it is only found in the outer regions of the PSS. 
An example of this kind of motion is shown in Fig.~\ref{fig:Braiding}.\par

\subsection{Evolution of the PSS for increasing values of the energy}
\begin{figure}[!h]
\begin{tabular}{cc}
\includegraphics[width = 0.49\textwidth]{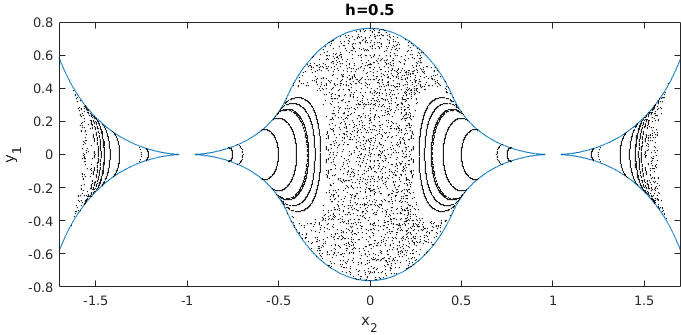}&\includegraphics[width = 0.49\textwidth]{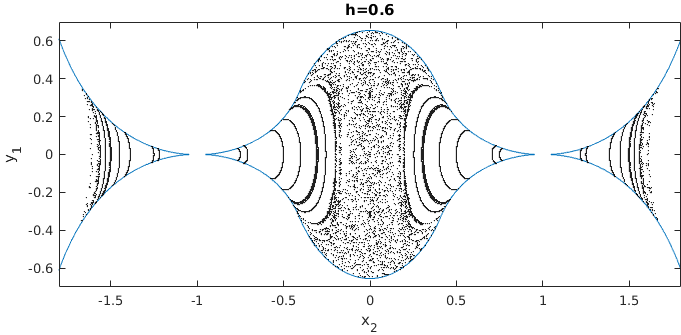}\\
\includegraphics[width = 0.49\textwidth]{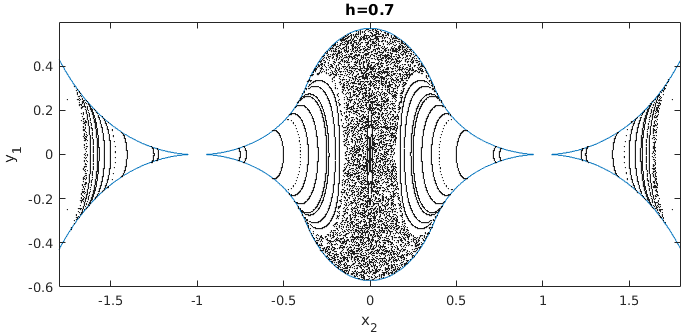}&\includegraphics[width = 0.49\textwidth]{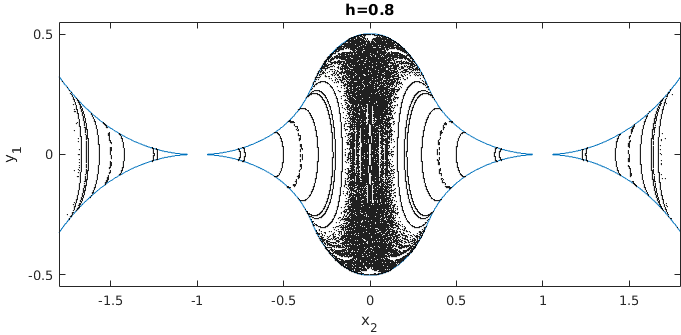}\\ 
\includegraphics[width = 0.49\textwidth]{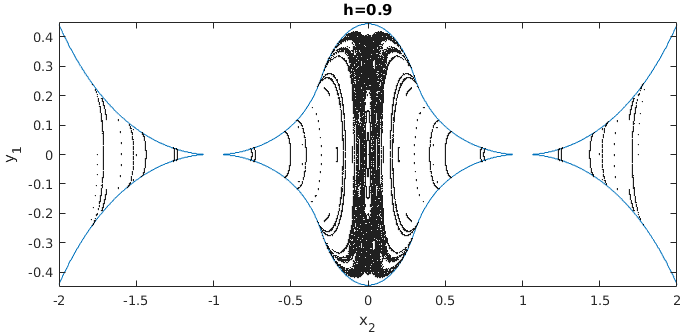}&\includegraphics[width = 0.49\textwidth]{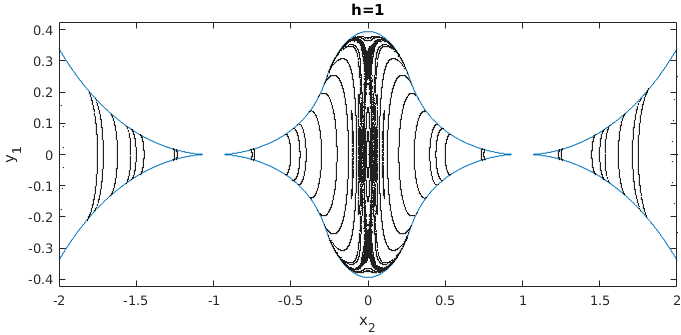}\\
 \includegraphics[width = 0.49\textwidth]{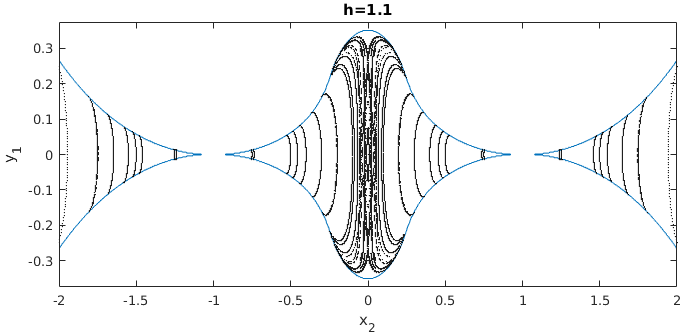}&\includegraphics[width = 0.49\textwidth]{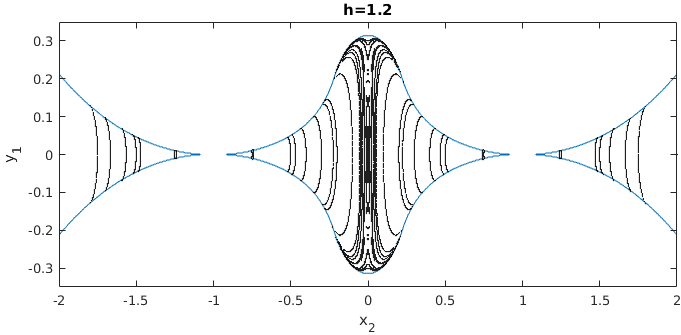} 
\end{tabular}
\caption{PSS for various values of $h$ showing the formation of the leapfrogging envelope and its growth as $h$ increases.}
\label{MultiSection}
\end{figure}

After categorizing the qualitatively different regions of motion in the PSS for a typical value of the energy of the system ($h=0.7458$) we construct the PSSs for increasing values of the energy, in order to study the evolution of the sections (Fig.~\ref{MultiSection}).

The first thing that someone can observe is that as the energy of the system increases, the permissible area of motion decreases. In addition, the central chaotic region also decreases and it is partially replaced by the leapfrogging envelope which appears for $h>0.6$ and gradually, as the value of the energy increases, it becomes more clearly discernible. For values $h>1$ the chaotic layer becomes so thin that the PSS consists practically only of regular motions. This complies with the fact that for low energies (which in this central region corresponds to small values of $\alpha$) where the vortex pairs are close and have similar diameters no leapfrogging configurations can be found. On the other hand, as the energy increases (and the value of $\alpha$ also increases) leapfrogging motion is possible and when $h$ increases more, in which case
the vortex pairs are well separated, then all the possible motions are regular. 
On the other hand, we see that the braiding and walkabout motion regions exist for every examined value of the energy, {although their particular boundaries move as $h$ varies}.

\subsection{Bifurcation analysis}\label{bifurcations}
\begin{figure}[!h]
        \centering
				\includegraphics[width = 0.7\linewidth]{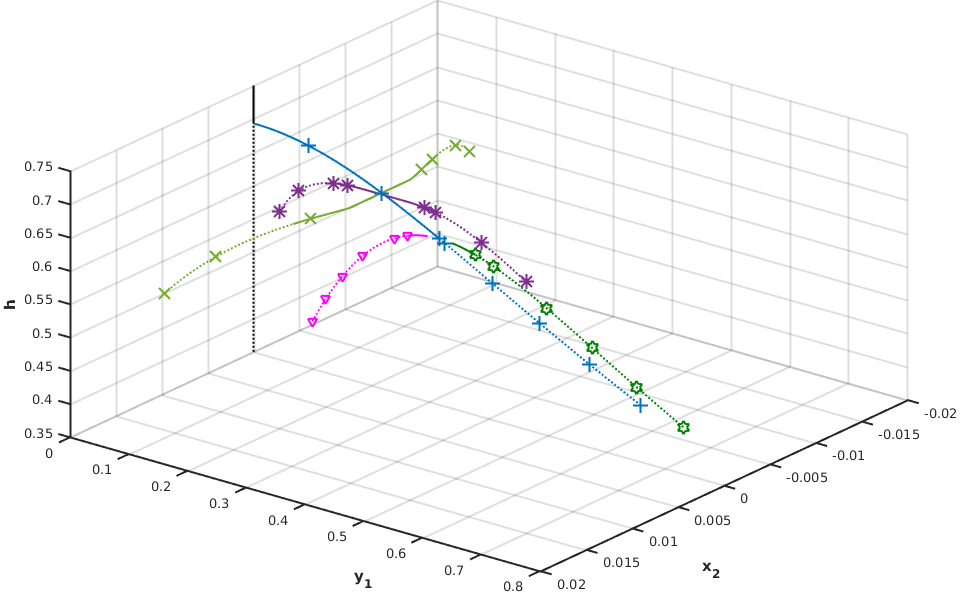}
                \includegraphics[width = 0.7\linewidth]{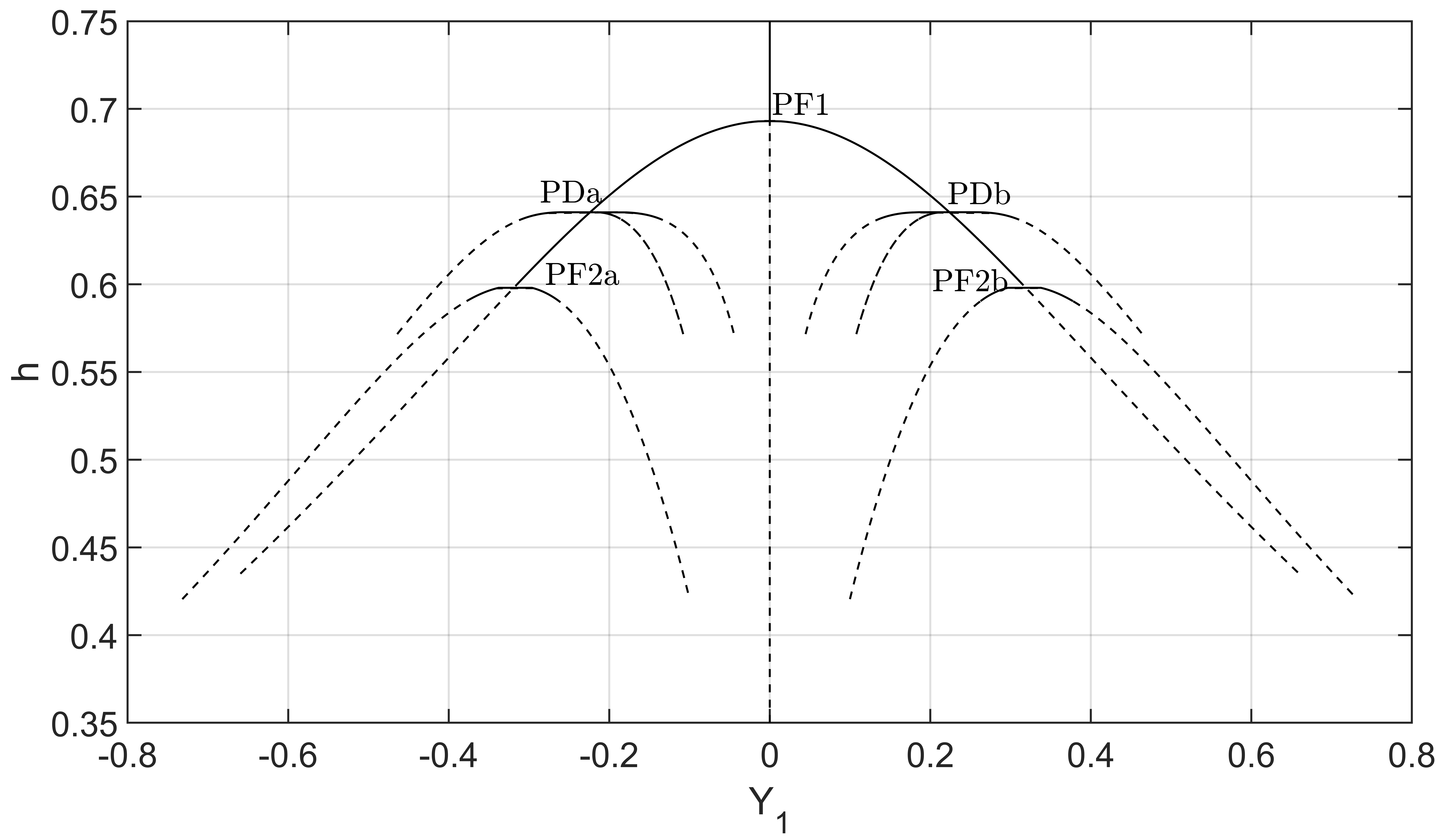}
                                \caption{In the upper panel of this figure the Bifurcation tree Diagram for the main leapfrogging solution. The main branch of this tree is the main leapfrogging solution with the bifurcated solutions spawning from it (distinguished by independent
                                  markers and colors). Only half of the tree is shown, the other half is the same but mirrored across the $y_1 = 0$ plane.
                                  Solid lines represent linearly stable motions while the dotted ones represent unstable motions.
                                  In the bottom panel of the figure, a two-dimensional projection of
                                  the diagram in the $(Y_1,h)$ plane
                                  is also given with an identification
                                  of the relevant bifurcations (PF stands
                                  for pitchfork and PD for period doubling).}
        \label{fig:Tree}
\end{figure}

We now proceed to the bifurcation analysis of the
leapfrogging periodic orbit.
A partial bifurcation tree is shown in Fig.~\ref{fig:Tree} 
. The branches of the tree each represent initial conditions for an orbit on the section, markers are used to distinguish between independent solutions. The main branch corresponds to the main leapfrogging solution. Starting from the top of the tree and moving downward, the first feature one notices is the supercritical pitchfork bifurcation at $ h \approx 0.6931$. Previous literature~\cite{Tophoj} has shown that the leapfrogging solution becomes unstable when the parameter $\alpha$ is below $ \phi^{2} \approx 0.3819$ where $\phi$ is the golden ratio. By using (\ref{eq:Alpha2Ham}) we find that these two values are in agreement, and conclude that this supercritical pitchfork is the explanation for the observed transition.
\begin{figure}[!h]
        \centering
				\includegraphics[width = 0.7\linewidth]{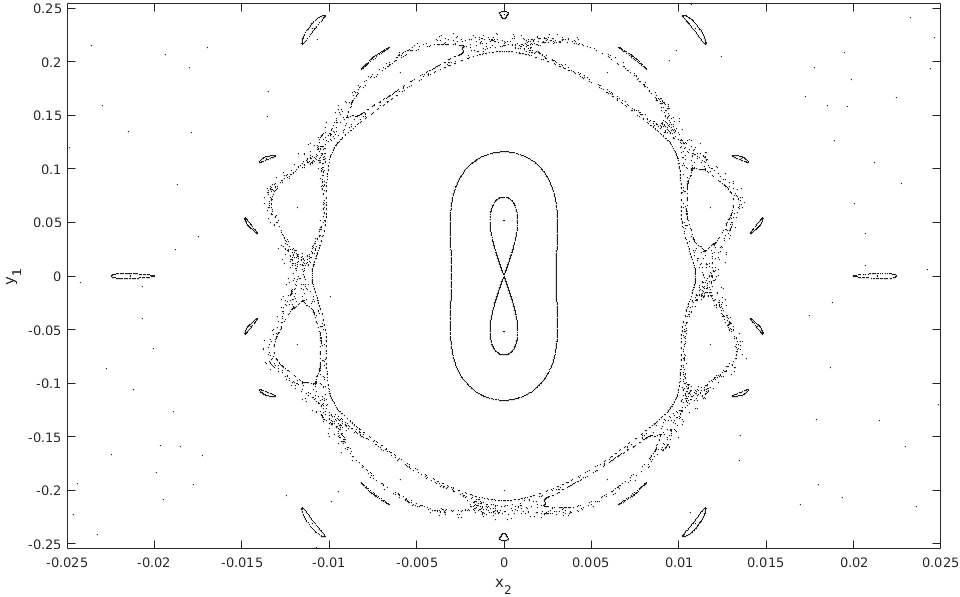}
                                \caption{Leapfrogging Envelope for $h=0.69$. This value lies below the stability threshold ($h\approx 0.6931$). We can see here that the central stable periodic orbit has been replaced by an unstable one
                                  (a saddle
                 point)
                                  while two new stable orbits have been created above and below the central one. We can still observe the higher resonances.}
        \label{fig:BifurcatedEnvelope}
\end{figure}

Fig.~\ref{fig:BifurcatedEnvelope} shows how the leapfrogging envelope has changed below the stability threshold. The central stable periodic orbits have been replaced by an unstable one (a saddle point) while the figure-eight-shaped invariant curve originating from the unstable orbit includes two new stable periodic orbits which result from the pitchfork bifurcation. We call these new stable periodic orbits  {\it asymmetric leapfrogging motions}.

\begin{figure}[!h]
        \centering
				\begin{tabular}{cc}
				(a)&(b)\\
				\includegraphics[width = 8cm]{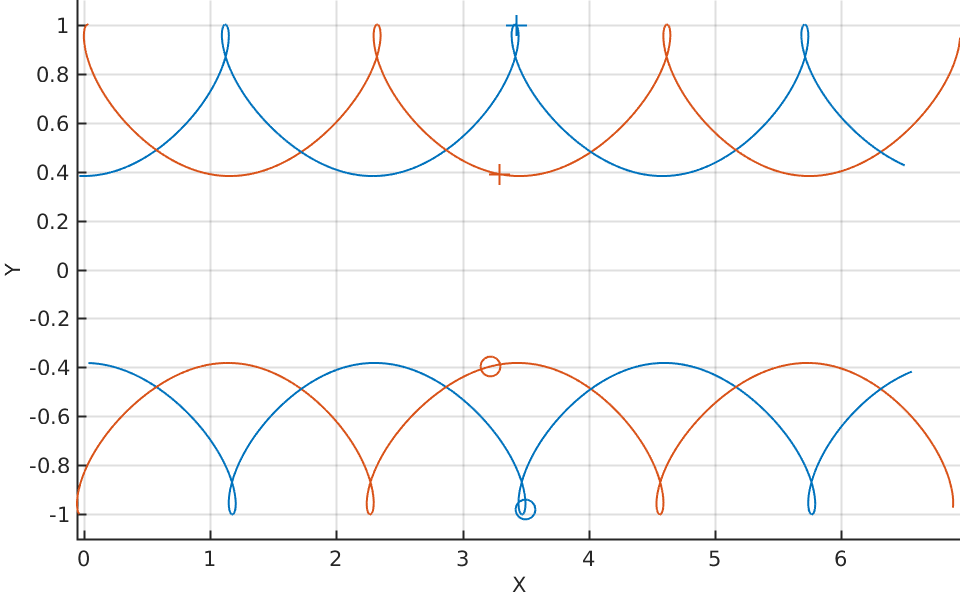} &\includegraphics[width = 8cm]{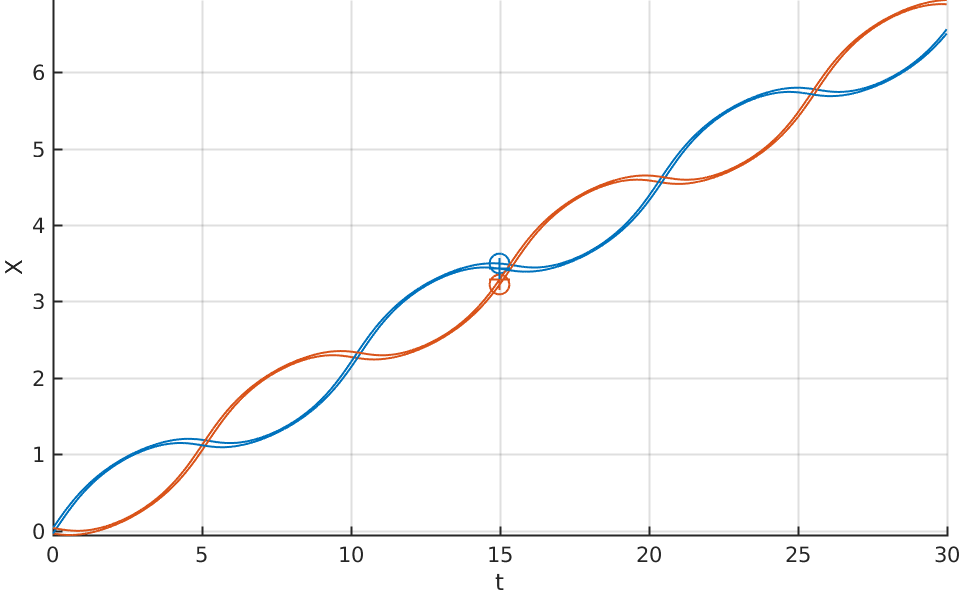}
				\end{tabular}
                                \caption{
Temporal evolution of the Asymmetric Leapfrogging orbit for $h=0.69$, $x_2=0$, $y_1=0.052$. (a): The motion in the $X$-$Y$ plane. (b): The motion in the $t$-$X$ plane. We can observe that one member of each pair is always lagging with respect to the other. Here and henceforth, by the same line color we denote the vortices of the same pair, while by the same symbol we denote vortices of the same circulation.}
        \label{fig:BifurcatedLeapfrogging}
\end{figure}

In Fig.~\ref{fig:BifurcatedLeapfrogging}(a) an example of this new kind of motion in the original space is shown. In this figure we can observe that the asymmetric leapfrogging motion is very similar to the original motion. The important distinction is that the pairs are not oriented precisely vertically in the asymmetric motion. This is made more obvious by Fig.~\ref{fig:BifurcatedLeapfrogging}(b),
where the temporal behavior of this solution is inspected;. 
In the previously known solution, the $X$-coordinates of both members of a single vortex pair were exactly the same. In this new motion, we see that the $X$-coordinates are never exactly the same and so one member of the pair lags behind the other. The lag varies at different points in the interval between section hits, but is always a constant when the orbit hits the section, since this motion is also periodic. \par

\begin{figure}[!h]
	\centering
	\includegraphics[width = 0.7\linewidth]{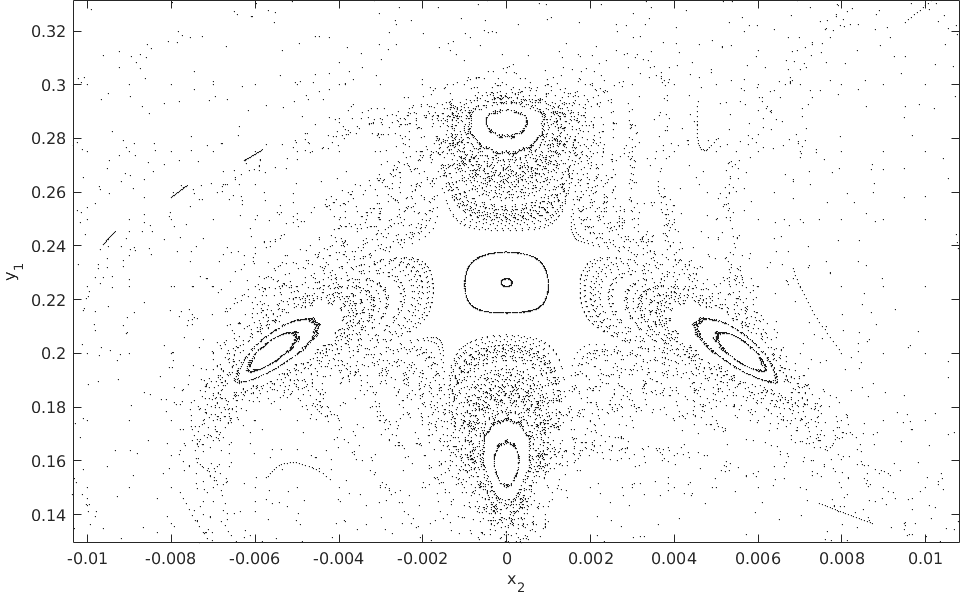}
	\caption{Close up view of a period-doubled region on the Poincar\`e section for $h=0.64$. The parent orbits remains stable after the period doubling, which generates two new period-two orbits which introduce four stability islands on the section.}
	\label{fig:PDSection}
\end{figure}

Continuing further down the bifurcation tree, the next event
--one that, along with the asymmetric solutions revealed above,
to the best of our knowledge, has not been identified previously--,
is a period doubling of the asymmetric leapfrogging orbits. Fig.~\ref{fig:PDSection} shows the period doubling on the Poincar\`e section for
one of the branches which occurs around $h=0.64$. This is a degenerate bifurcation which generates two
linearly stable period-two orbits off each branch, while the parent orbit
remains also stable resulting in a total of four new periodic orbits.

In this case, a pair of Floquet multipliers $\lambda$ meets at $\lambda=-1$ and instead of exiting the unit circle along the real axis, as would normally be the case for a period doubling, they cross each other and
remain on the unit circle. This degenerate scenario of period doubling
is consequently responsible for the birth of two separate period doubled
solutions, as shown in Fig.~\ref{fig:PDSection}.

\begin{figure}[!h]
\centering
\includegraphics[width = 10cm]{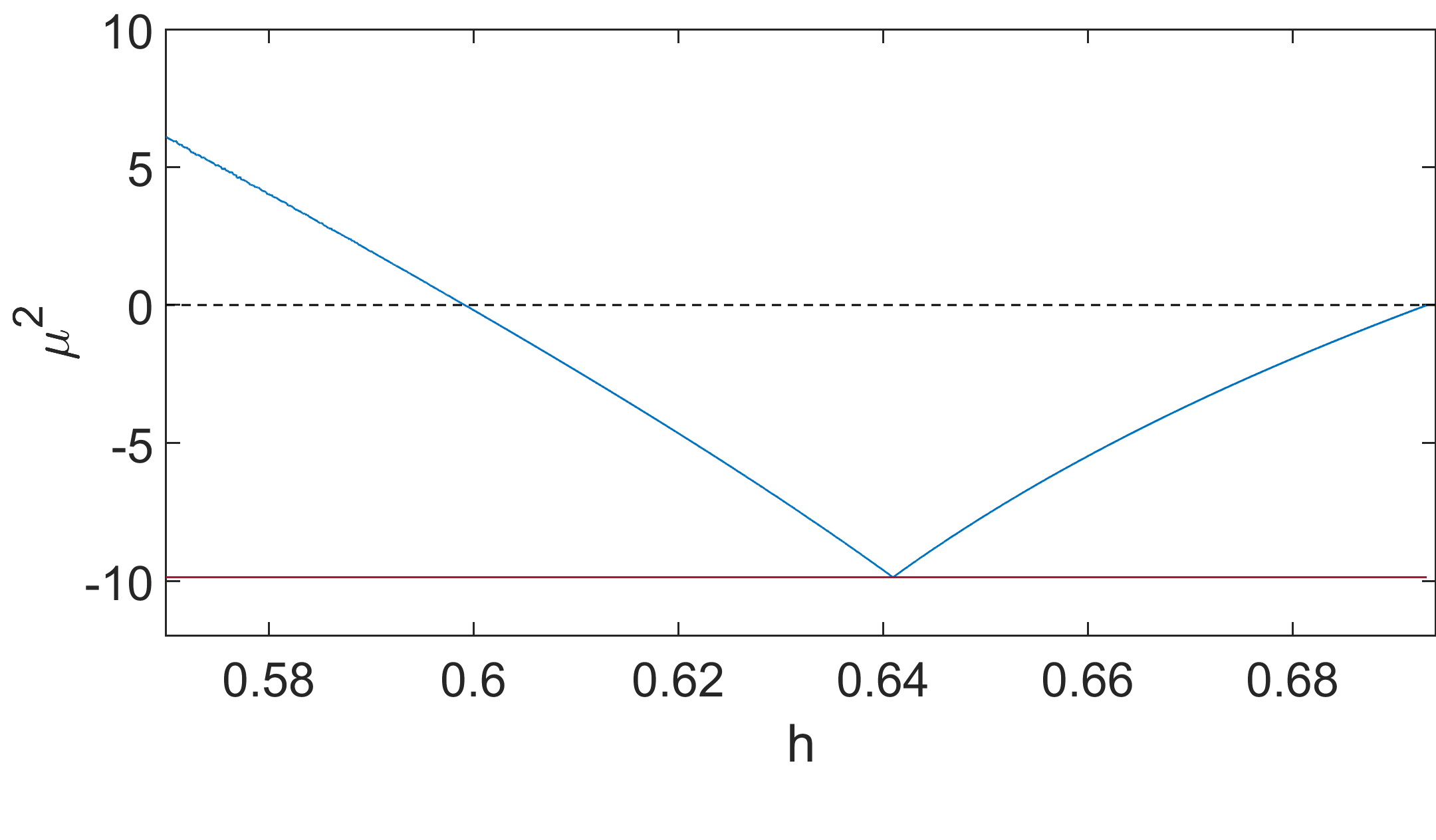}
\caption{A plot of the square of the normalized Floquet exponent of the asymmetric leapfrogging mode. It acquires the value of $\mu^2=-\pi^2$ around $h=0.641$ (where the period doubling occurs), while it remains negative for the range $0.5988<h<0.6931$ indicating the linear stability of the orbit. The value of $\mu^2$ becomes positive below $h=0.5988$ where a pitchfork bifurcation occurs and the branch becomes unstable.}
\label{floquet}
\end{figure}

This can be shown more clearly in Fig.~\ref{floquet}, where the sqare of the corresponding normalized Floquet exponents, which are defined through $\mu=\ln(\lambda)$, is shown. In this figure we follow the value of $\mu^2$ for a stable asymmetric leapfrogging branch of the pitchfork bifurcation occurring for $h=0.6931$. It remains negative, which means $\mu \in \mathbb{I}$ thus, it corresponds to linearly stable motion. For $h=0.641$, $\mu^2$ takes the value $-\pi^2$ which corresponds to $\lambda=-1$ and the period doubling bifurcation occurs. The orbit remains stable up to $h=0.5988$. 

\begin{figure}[!h]
        \centering
				\begin{tabular}{cc}
				(a)&(b)\\
				\includegraphics[width = 8cm]{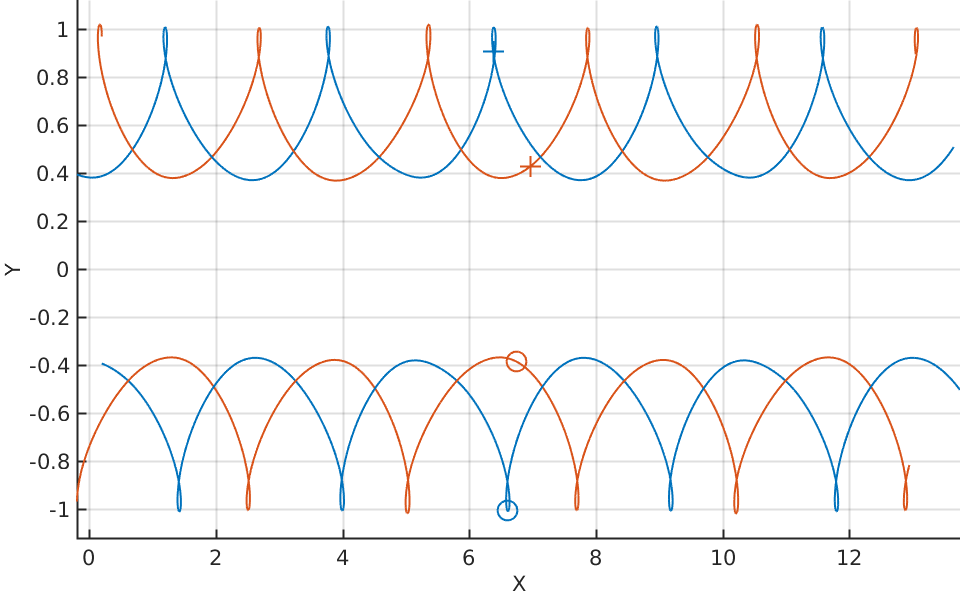}&\includegraphics[width = 8cm]{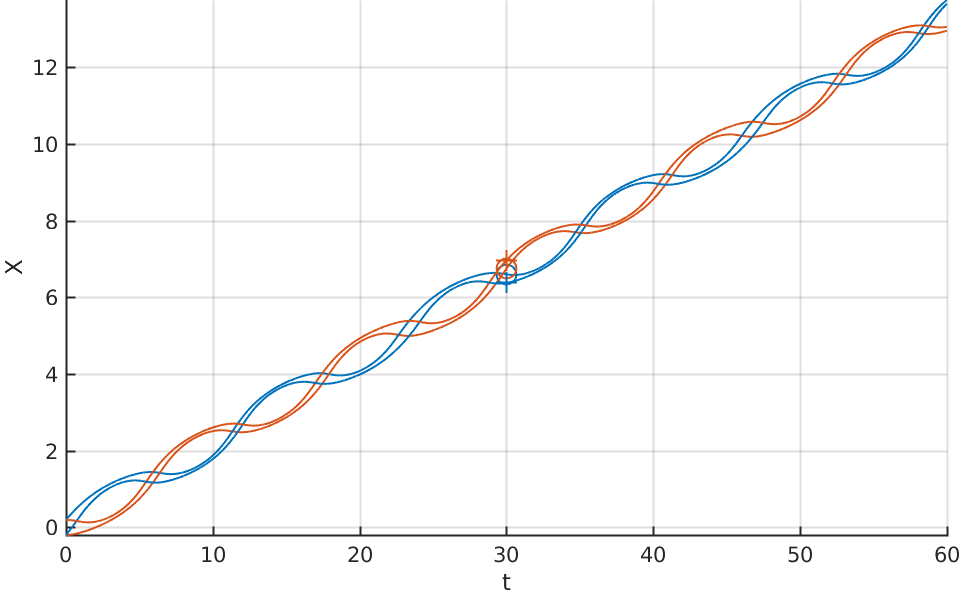}
				\end{tabular}
        \caption{Temporal evolution of the Period-Doubled leapfrogging with $h = 0.64$. (a): The motion in the $X$-$Y$ plane. (b): The motion in the $t$-$X$ plane.  We can see here that the lag between the two vortex pairs takes two periods of leapfrogging to return to its initial state.}
        \label{fig:PDLeapfrog}
\end{figure}

Fig.~\ref{fig:PDLeapfrog} shows the temporal evolution of the new orbits. Comparing \ref{fig:PDLeapfrog}(b) to \ref{fig:BifurcatedLeapfrogging}(b) reveals that the lag in the period doubled orbits varies periodically at the section hits. This is the additional quality that these orbits possess which gives them  a higher resonance. In Fig.~\ref{fig:PDLeapfrog}(b) we see that the lag takes two periods of leapfrogging to return to its initial state hence the two different section hit points.

\begin{figure}[!h]
	\centering
	\includegraphics[width = .7\linewidth]{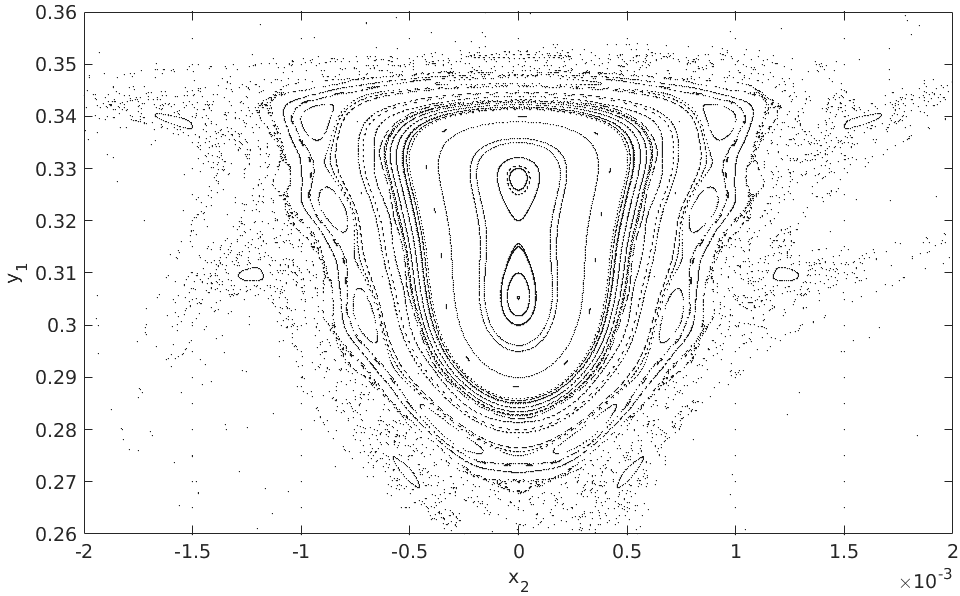}
	\caption{Section in the region of one of the branches after undergoing a second pitchfork bifurcation, $h=0.5988$. We can see that the central stable periodic orbit has been replaced by an unstable one (a saddle point), while two new stable
          more asymmetric orbits has been generated above and below
          the original one.}
	\label{Pitchfork2}
\end{figure}

\begin{figure}[!h]
	\centering
	\begin{tabular}{cc}
	(a)&(b)\\
	\includegraphics[width = 8cm]{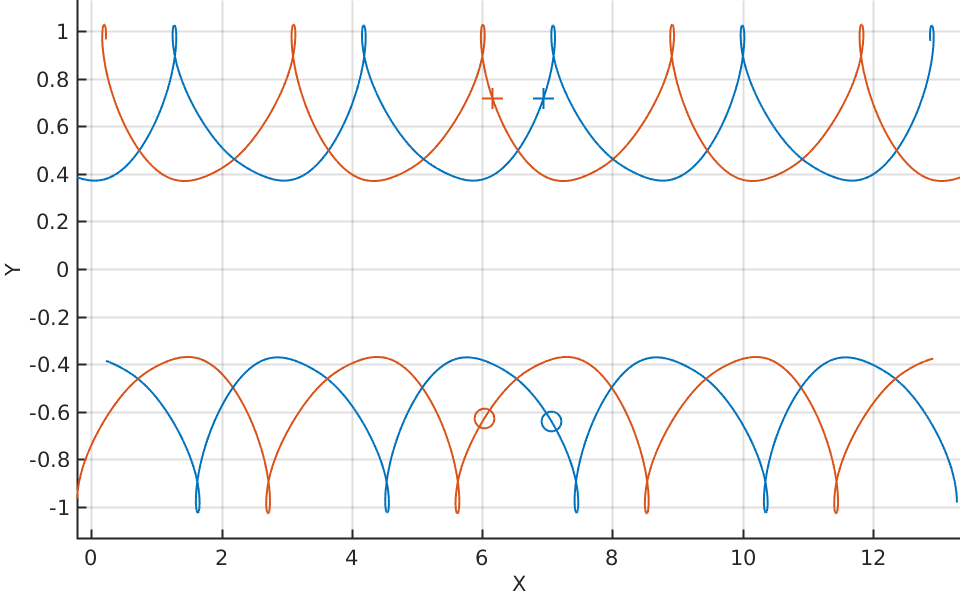}&\includegraphics[width = 8cm]{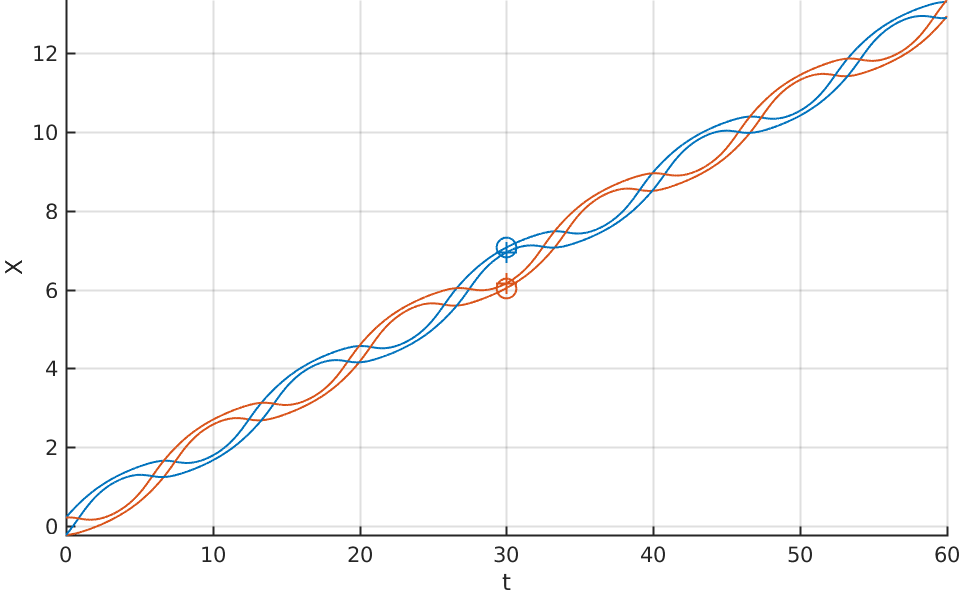}
	\end{tabular}
	\caption{Temporal evolution of the orbit resulting from the second pitchfork bifurcation. (a): The motion in the $X$-$Y$ plane. (b): The motion in the $t$-$X$ plane. Now, the characteristic lag between the two vortex-pairs is again constant at all of the section hits. The magnitude of the lag however has gotten much bigger than the case of Fig.~\ref{fig:BifurcatedLeapfrogging}(b), and we see much more apparent asymmetries than in the parent branch.}
	\label{Pitchfork2Leapfrogging}
\end{figure}

Turning back to the parent branch of the period doubled motion, we continue downward to find another supercritical pitchfork bifurcation which occurs for $h=0.5988$ (as it can be seen also in Fig.~\ref{floquet}). Fig.~\ref{Pitchfork2} shows the region on the section where the bifurcation occurs for one of the branches. The typical picture of a pitchfork bifurcation is again depicted in this figure like in Fig.~\ref{fig:BifurcatedEnvelope}. The panels of Fig.~\ref{Pitchfork2Leapfrogging} show the temporal evolution of the new motions. These motions show the characteristic lag that the parent branch exhibits, and the lag is once again constant at all of the section hits. The magnitude of the lag, however, has gotten much bigger (in comparison,
e.g., to Fig.~\ref{fig:BifurcatedLeapfrogging}) and we see much more
evident asymmetries than in the parent branch.  Higher order resonances
are still discernible in this case as well.

\begin{figure}[!h]
\begin{tabular}{cc}
(a)&(b)\\
	\includegraphics[width = 8cm]{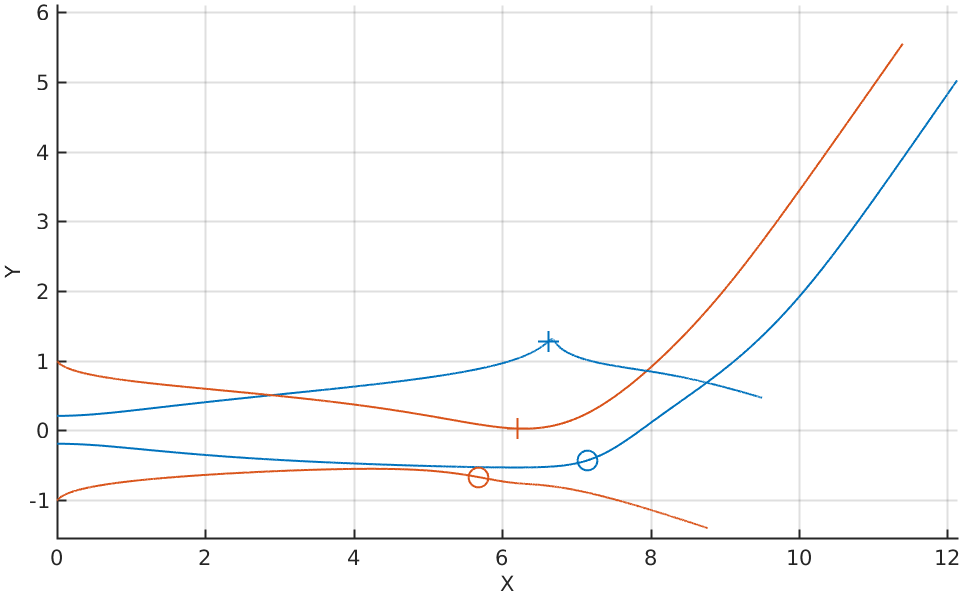}&\includegraphics[width = 8cm]{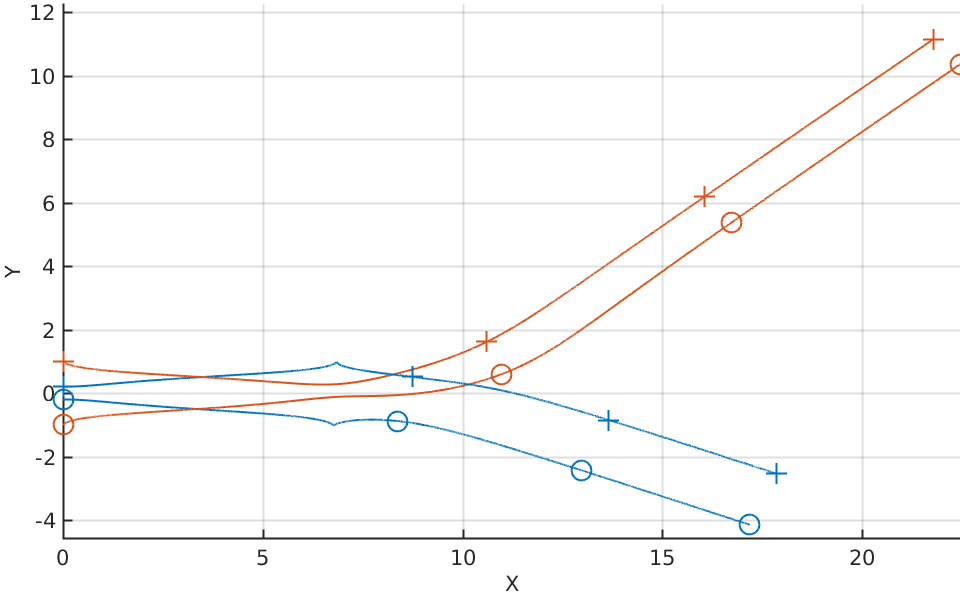}
\end{tabular}
\caption{(a): A similar example to the instability observed in~\cite{Acheson}. (b): Another orbit that undergoes a variant
  of this instability (with a different outcome). Both orbits shown are for a value of $h=0.1116$. (a) has $(x_2, y_1)=(0.001, 0)$ and (b) has $(x_2, y_1)=(-0.01, 0.001)$.}
	\label{Disintegration}
\end{figure}

The sequence of pitchfork bifurcations that we have already
analyzed suggests that as the energy continues to decrease, the region will continue to undergo more and more bifurcations in a cascade like pattern. This cascade continues until the entire leapfrogging envelope has practically disappeared as it can be seen in Fig.~\ref{MultiSection} for $h<0.7$.
However, the nature of the instabilities of this system as discussed in
both~\cite{Acheson} and in~\cite{Tophoj}
make it quite difficult to identify the details of the relevant bifurcation
at lower energies. This is because the primary instability, carries orbits off the section in such a way that they never return. Fig.~\ref{Disintegration} shows examples of orbits that are affected by this instability. Fig. \ref{Disintegration}(a) demonstrates the case which was studied in~\cite{Acheson}. Fig. \ref{Disintegration}(b) shows a case which is very similar, the primary difference being that in \ref{Disintegration}(a) the pairs split and exchange partners, whereas in \ref{Disintegration}(b) the pairs are preserved.

\section{Conclusions}
\label{conclusions}

In this work we studied the dynamic interaction between two vortex pairs. In particular we revisited the classic theme (dating back
to the works of Helmholtz and Love~\cite{Love}) of leapfrogging and leapfrogging-like motions. Earlier  studies had focused on exploring this under a dynamic
formulation using direct numerical simulations~\cite{Acheson} and
analysis of the associated (effective) periodic orbit~\cite{Tophoj}.
Here, we developed a Hamiltonian viewpoint of the associated
phase space, enabling us to acquire a broader perspective on the
possible motions therein, as well as the orbits related to leapfrogging
(generalized leapfrogging, walkabout, braiding motion, asymmetric orbits,
period doubled variants thereof, etc.).

Under the appropriate transformations the system of interest
was transformed into a two-degrees of freedom one. On the basis of
this reduction, we constructed a Poincar\'e Surface of Section for a typical
energy of the system in order to categorize the various types of motion of
the system, the main of them being the leapfrogging, braiding and walkabout ones. After that we produced a series of PSS for increasing values of the energy to observe that most of the regions of motion exist in all of the PSS. A significant difference that occurs between the sections is that the leapfrogging envelope disappears for decreasing values of the energy and is replaced my a chaotic region. This fact plays a crucial role in the overall stability of the leapfrogging motion. 

Finally, we focused our attention on
the central leapfrogging solution and produced in a systematic way
(at least in as far as the first few branches are concerned) its bifurcation
tree. We used the total energy of the system as a parameter, providing
a systematic explanation of the existence of the stability thresholds
mentioned in earlier works. The landscape of resulting bifurcations
(pitchfork cascades, degenerate period doublings, etc.) manifested
a wealth of phenomena and additional coherent motions that were not
previously revealed, to the best of our knowledge.

This effort paves the way for numerous possible paths in the future
for related systems. While the wealth of associated phenomena
may grow considerably and the insight of the techniques used herein
is more limited to low-dimensional settings, it would certainly
be worthwhile to examine the possibility of leapfrogging phenomena/features
in settings involving 6- or 8- vortices, in the spirit of the computations
of~\cite{wacks} (and associated experiments discussed therein) for
generalized leapfrogging involving more vortex pairs.
For the 3-dimensional realm of vortex lines and vortex rings,
studying in a systematic way the
phase plane and possible motions both along the lines of~\cite{konstantinov}
in the absence of a trap,
as well the variant in the presence of the trap~\cite{wang2} will be
of particular interest as a direction for future work.

\vspace{10mm}

{\bf Acknowledgements}: PGK gratefully acknowledges support from
NSF-PHY-1602994. VK and PGK where also supported by 
the ERC under FP7; Marie Curie Actions, People, International Research
Staff Exchange Scheme (IRSES-605096).

\bibliographystyle{plain} 
\bibliography{./Bibliography_v3}

\end{document}